\documentclass[11pt,a4paper]{amsart}
\usepackage{amsmath}
\usepackage{amssymb}
\usepackage{amsfonts}
\usepackage{hyperref}

\def\={\ =\ }

\theoremstyle{plain}

\numberwithin{equation}{section}

\begin{document}
\title[Chern-Simons theory as a 2d Yang-Mills theory\hfill\ \ ]{Unitary
Chern-Simons matrix model and the Villain lattice action}
\urladdr{}
\author{Mauricio Romo}
\address{\flushleft Department of Physics \\
University of California Santa Barbara \\
California 93106, USA}
\email{mromo@physics.ucsb.edu}
\author{Miguel Tierz}
\address{\flushleft Departamento de An\'{a}lisis Matem\'{a}tico, Facultad de
Ciencias Matem\'{a}ticas.\\
Universidad Complutense de Madrid. \\
Plaza de Ciencias 3, 28040, Madrid. Spain.}
\email{tierz@mat.ucm.es}
\address{\flushleft Grupo de F\'{\i}sica Matem\'{a}tica\\
Complexo Interdisciplinar da Universidade de Lisboa \\
Avenida Professor Gama Pinto, 2, PT-1649-003 Lisboa, Portugal.}
\email{tierz@cii.fc.ul.pt}
\urladdr{}
\curraddr{ }

\begin{abstract}
We use the Villain approximation to show that the Gross-Witten model, in the
weak- and strong-coupling limits, is related to the unitary matrix model
that describes $U(N)$ Chern-Simons theory on $S^{3}$. The weak-coupling
limit corresponds to the $q\rightarrow 1$ limit of the Chern-Simons theory
while the strong-coupling regime is related to the $q\rightarrow 0$ limit.
In the latter case, there is a logarithmic relationship between the
respective coupling constants. We also show how the Chern-Simons matrix
model arises by considering two-dimensional Yang-Mills theory with the
Villain action. This leads to a $U(1)^{N}$ theory which is the
Abelianization of 2d Yang-Mills theory with the heat-kernel lattice action.
In addition, we show that the character expansion of the Villain lattice
action gives the q deformation of the heat kernel as it appears in $q$%
-deformed 2d Yang-Mills theory. We also study the relationship between the
unitary and Hermitian Chern-Simons matrix models and the rotation of the
integration contour in the corresponding integrals.
\end{abstract}

\maketitle

\section{Introduction}

In the 1970s, Wilson introduced and studied lattice versions of Yang-Mills
(YM) theory, in which the dynamical variables are elements of the gauge
group, defined on the links that connect the adjacent lattice points \cite%
{Wilson:1974sk}. The choice of the lattice action in Yang-Mills theory was
an important aspect of the problems considered in the early development of
lattice gauge theory \cite{Wilson:1974sk}-\cite{Onofri:1981qk}. In
particular, in the late 1970s and early 1980s, a number of papers studied
this possibility, leading to the consideration of several alternatives to
the Wilson action \cite{Wilson:1974sk}, like the heat-kernel action \cite%
{Susskind:1979up,Stone:1978pe,Menotti:1981ry} or the Manton action \cite%
{Manton:1980ts}. One of the motivations was the need for a proper
understanding of the transition from strong to weak coupling in lattice
gauge theories.

A well-known result in the study of non-Abelian two-dimensional Yang-Mills
theory with the Wilson action is the third-order phase transition, found by
Gross and Witten \cite{GW} and Wadia \cite{Wadia:1980cp}, in a one-plaquette
model, described by a unitary matrix model. This result has turned out to be
of relevance in many current problems in theoretical physics, like the study
of Hagedorn and deconfining transitions in weakly coupled Yang-Mills theory 
\cite{Aharony:2003sx}. The Gross-Witten model has been also recently
discussed in the study of type 0B and 0A fermionic string theories \cite{Kms}
and in relation with other solvable models, like the Kontsevich model \cite%
{Morozov:2009jv}.

We will first show that the unitary matrix model of $U(N)$\ Chern-Simons
theory on $S^{3}$ \cite{Okuda:2004mb,Szabo:2010sd,Ooguri:2010yk} is
intimately related to the Gross-Witten model when one considers it together
with the Villain approximation of the XY model \cite%
{Villain:1974ir,Jose:1977gm}. We introduce the Villain approximation,
together with the two relevant matrix models, in the next section. We will
show that both the weak-coupling and strong-coupling regimes of the
Gross-Witten model can be described, using the Villain approximation, by
analytically continued $U(N)$ Chern-Simons on $S^{3}.$ Recall that the
Chern-Simons action is given by \cite{Witten:1988hf}%
\begin{equation}
S_{\mathrm{CS}}(A)={\frac{k}{4\pi }}\int_{M}\mathrm{Tr}(A\wedge dA+{\frac{2}{%
3}}A\wedge A\wedge A),  \label{cs}
\end{equation}%
where $A$ is the connection, a 1-form valued on the corresponding Lie
algebra, and $k\in \mathbb{Z}$ is the level. The $q$-parameter is defined in
terms of the level $k$ by $q=\exp \left( 2\pi i/(k+N)\right) $.

We will see that the weak-coupling limit corresponds to the $q\rightarrow 1$
limit of the Chern-Simons theory, whereas the strong-coupling limit of the
Gross-Witten model corresponds to the opposite limit, $q\rightarrow 0$. In
the matrix model formulation, the $q$ parameter is treated as real and
written in terms of a coupling constant $g_{s\text{ }}$as $q=\mathrm{e}%
^{-g_{s}}$ \cite{Tierz:2002jj}. The above characterization of Chern-Simons
theory as being analytically continued precisely refers to this treatment of 
$q$ as a real parameter.

Indeed, actual computations with the matrix model are carried out with $q$
real, using for example the associated $q$-orthogonal polynomials \cite%
{Tierz:2002jj}, and the identification $g_{s}=2\pi i/(k+N)$ at the end,
allows us to make contact with the well-known expressions for the
Chern-Simons observables \cite{Witten:1988hf}. See for example \cite%
{Tierz:2002jj}, where the simple case of the $U(N)$ Chern-Simons partition
function on $S^{3}$ is computed with the Stieltjes-Wigert polynomials with a 
$q$ parameter, $q=\mathrm{e}^{-g_{s}}$.

The correspondence between the Gross-Witten model and the Chern-Simons
matrix model is of a rather different nature in the two opposite limits $%
q\rightarrow 0$ and $q\rightarrow 1$. In the weak-coupling limit $g_{\mathrm{%
\ YM}}^{2}\rightarrow 0$, as shown already in \cite{Szabo:2010qv}, the
coupling constants are related by $g_{\mathrm{YM}}^{2}=2g_{s},$ whereas in
the strong-coupling limit we will have that $g_{\mathrm{s}}=2\ln \left( g_{%
\mathrm{YM}}^{2}\right) $ or, equivalently, $q=1/g_{\mathrm{YM}}^{4},$ as we
shall see in Sec. 2 in detail.

We will end Sec. 2 by exploring some consequences of these relationships
between the models. In particular, in Sec. 2.1 we show that the Gross-Witten
model at weak coupling is a Gaussian matrix model whose free energy has an
expansion that can be interpreted in terms of closed strings.

The relationship between the Gross-Witten model and the Chern-Simons matrix
model, based on the application of the Villain approximation to the Wilson
action, indicates that the direct consideration of the Abelian Villain
action in lattice two-dimensional Yang-Mills theory, should describe
Chern-Simons theory. Two-dimensional Yang-Mills theory was also studied with
the Manton action \cite{Lang:1980sz,Lang:1980ws} and the heat-kernel action 
\cite{Menotti:1981ry}. We shall see that, indeed, the straightforward
generalization of the Abelian $U(1)$ lattice action, which is just a theta
function \cite{Villain:1974ir,Banks:1977cc,Peskin:1977kp}%
\begin{equation}
\exp (-S_{\mathrm{V}}\left( \phi \right) )=\sum_{l=-\infty }^{\infty }%
\mathrm{e}^{-\frac{1}{g^{2}}\left( \phi +2\pi l\right) ^{2}},
\label{A-Villain}
\end{equation}%
to the non-Abelian case, in the setting of $U(N)$ two-dimensional Yang-Mills
theory, directly gives $U(N)$ Chern-Simons theory on $S^{3}$. This
straightforward extension of the Villain lattice action to the non-Abelian
case was explored by Onofri, shortly after the study of the heat-kernel case 
\cite{Menotti:1981ry}, in a less well-known work \cite{Onofri:1981qk}. The
description of pure Chern-Simons theory by such a model has not hitherto
been realized.

It is well known that the Villain model arises in the Kogut-Susskind
Hamiltonian lattice gauge theory \cite{Kogut:1974ag} in the Abelian case,
which leads to a direct correspondence with the planar Heisenberg (or XY)
model \cite{Polyakov:1978vu,Susskind:1979up}. The non-Abelian case leads to
the heat kernel \cite{Susskind:1979up,Menotti:1981ry}, and we shall see that
the Chern-Simons matrix model follows from Abelianization of the heat-kernel
propagator in the context of two-dimensional Yang-Mills theory. This is the
content of Sec. 3 and, in particular, we show in Sec. 3.1 that this Abelian
projection is equivalent to a $q$ deformation of 2d Yang-Mills, in
consistency with \ the known relationship between Chern-Simons theory and a $%
q$ deformation of 2d Yang-Mills theory \cite{Aganagic:2004js}. Recall that
Chern-Simons theory is known to be explained in terms of an Abelian
two-dimensional Yang-Mills theory, as was shown at the level of the path
integral, first in the case of manifolds of the type $S^{1}\times \Sigma _{h}
$, where $\Sigma _{h}$ denotes a Riemann surface of genus $h$ \cite%
{Blau:1993tv} and, more recently, for Seifert fibrations over $\Sigma _{h}$ 
\cite{Blau:2006gh}, which contains the $S^{3}$ case, the one studied in this
paper at the level of the matrix model.

To conclude, we study in the Appendix the precise relationship between the
unitary and the Hermitian versions of the Chern-Simons matrix model focusing
also in the rotation of the contours of integration.\newline

\section{Gross-Witten model and the Villain approximation}

The approximation devised by Villain in 1975 in the study of the
two-dimensional XY\ model \cite{Villain:1974ir} is based on the simple
observation that the term $\exp \left( \beta \cos \theta \right) $ that
appears in the 2d XY model can be well approximated for large $\beta $ by a
periodic Gaussian with minima in the same locations and with the same
curvature. That is 
\begin{equation}
\exp \left( \beta \cos \theta \right) \sim \mathrm{e}^{\beta
}\sum_{n=-\infty }^{\infty }\mathrm{e}^{-\frac{1}{2}\beta \left( \theta
-2\pi n\right) ^{2}}\text{ for }\beta \rightarrow \infty .  \label{Villain}
\end{equation}%
But the l.h.s. term is of course also the weight function of the matrix
model description of the one-plaquette model of Yang-Mills theory based on
the Wilson action (namely, the Gross-Witten model \cite{GW}). The r.h.s is a
theta function and then the Villain approximation applied to the
Gross-Witten model leads to the relationship with a unitary matrix model
with a theta function as weight function.

Precisely, the unitary matrix model that describes $U(N)$ Chern-Simons
theory on $S^{3\text{ }}$ \cite{Okuda:2004mb,Szabo:2010sd,Ooguri:2010yk} is
given by \footnote{%
The unitary matrix model (\ref{UCS}) also describes Chern-Simons theory if
the weight function is $\Theta ^{-1}({\,-\mathrm{e}}\,^{{\,\mathrm{i}\,}%
\theta _{j}}|q)$ \cite{Szabo:2010sd}. This possibility has also been noticed
in \cite{Ooguri:2010yk}. The nonuniqueness description of the Chern-Simons
matrix models is described in \cite{Tierz:2002jj}. See the Appendix for its
precise relationship with the Hermitian matrix model} 
\begin{equation}
Z_{\mathrm{CS}}^{U(N)}\left( S^{3}\right) =\int_{0}^{2\pi
}\prod\limits_{j=1}^{N}\,\frac{\mathrm{d}\theta _{j}}{2\pi }~\Theta ({\,%
\mathrm{e}}\,^{{\,\mathrm{i}\,}\theta _{j}}|q)~\prod\limits_{k<l}\,\big\vert{%
\,\mathrm{e}}\,^{{\,\mathrm{i}\,}\theta _{k}}-{\,\mathrm{e}}\,^{{\,\mathrm{i}%
\,}\theta _{l}}\big\vert^{2},  \label{UCS}
\end{equation}%
where the weight function of the matrix model is a Jacobi third theta
function%
\begin{equation}
\omega \left( \theta \right) =\Theta ({\,\mathrm{e}}\,^{{\,\mathrm{i}\,}%
\theta _{j}}|q)=\sum_{n=-\infty }^{\infty }\mathrm{q}^{n^{2}/2}\mathrm{e}%
^{in\theta }.  \label{weight}
\end{equation}%
We show now how this model follows from the Gross-Witten model \cite%
{GW,Wadia:1980cp}, by using the Villain approximation \cite{Jose:1977gm}.
Aspects of the relationship between the two models, especially in the
weak-coupling regime, have already been studied in \cite{Szabo:2010qv},
using orthogonal polynomials. The results of the seminal works \cite%
{Villain:1974ir,Jose:1977gm} also allow us to extend the relationship
between the Gross-Witten and the Chern-Simons models to the strong-coupling
regime.

Recall that the Gross-Witten model is a unitary one-matrix model which
arises as the one-plaquette reduction of the combinatorial quantization of
Yang-Mills theory. In two dimensions the reduction is exact and described by
the partition function~\cite{GW} 
\begin{align}
Z_{N}^{\mathrm{GW}}(\beta )& :=\int_{U(N)}\,\mathrm{d}U~{\exp }\left( \frac{%
\beta }{2}\,\mathrm{Tr}\big(U+U^{\dag }\big)\right)   \label{GW} \\[4pt]
& =\int_{0}^{2\pi }~\prod\limits_{i=1}^{N}\,\mathrm{d}\theta _{i}~\mathrm{e}%
^{\beta \,\cos \theta _{i}}~\prod\limits_{i<j}\,\sin ^{2}\left( \frac{\theta
_{i}-\theta _{j}}{2}\right) \ ,  \notag
\end{align}%
where $\mathrm{d}U$ denotes the bi-invariant Haar measure for integration
over the unitary group $U(N)$. The $\beta $ parameter is usually written in
terms of the gauge coupling constant as $\beta =2/g_{\mathrm{YM}}^{2}$ \cite%
{GW} and hence, the strong coupling limit is given by $\beta \rightarrow 0$
and the weak coupling limit by $\beta \rightarrow \infty $.

The planar or XY model is characterized by a coupling between
nearest-neighbor spins which has the same analytical form as the potential
in the Gross-Witten model \cite{Villain:1974ir,Jose:1977gm} 
\begin{equation}
V_{\beta }=-\beta \left[ 1-\cos \left( \theta -\theta ^{\prime }\right) %
\right] .  \label{full-V}
\end{equation}%
The coefficients of the Fourier expansion of this potential are given by%
\begin{equation}
\mathrm{e}^{\widetilde{V}(s)}=I_{s}\left( \beta \right) ,  \label{Bessel}
\end{equation}%
where $I_{n}(z)$ is the modified Bessel function of order $n$. The partition
function of the Gross-Witten model can also be written as the determinant of
a Toeplitz matrix with (\ref{Bessel}) as entries of the matrix \cite%
{Bars:1979xb}%
\begin{equation}
Z_{N}(\beta )=\det_{1\leq i,j\leq N}\,\big[I_{i-j}(2\beta )\big]\ .
\label{IT}
\end{equation}%
On the other hand, for $U(N)$\ Chern-Simons theory on $S^{3}$ the
corresponding Toeplitz determinant is \cite{Szabo:2010sd}\footnote{%
Such a determinant was already considered in \cite{Onofri:1981qk}. See Sec.
3.}%
\begin{equation}
Z_{\mathrm{CS}}^{U(N)}\left( S^{3}\right) =\det_{1\leq i,j\leq N}\,\big[%
a_{i-j}(q)\big]\ ,\qquad \text{ with }\quad a_{j}=q^{{j^{2}}/{2}}\ .
\label{D}
\end{equation}%
In the weak-coupling limit $\beta \rightarrow \infty ,$ the Fourier
coefficient is%
\begin{equation}
\lim_{\beta \rightarrow \infty }\mathrm{e}^{\widetilde{V}(s)}=\mathrm{e}%
^{-s^{2}/2\beta }=\mathrm{e}^{-s^{2}g_{\mathrm{YM}}^{2}/4}.  \label{Gaussian}
\end{equation}%
In this limit, the Toeplitz determinant (\ref{IT}) has the Gaussian
coefficients (\ref{Gaussian}) as entries, and hence it coincides with (\ref%
{D}). Of course, $g_{\mathrm{YM}}^{2}\rightarrow 0$ in (\ref{Gaussian}) and
this implies $g_{s}\rightarrow 0$ on the Chern-Simons theory side as well.

The prescription in \cite{Villain:1974ir}, namely (\ref{Villain}), is valid
for both the opposite $\beta \rightarrow 0$ and $\beta \rightarrow \infty $
limits, using always periodic Gaussians, but including a renormalization
scale $R_{V}\left( \beta \right) $ and a rescaled inverse temperature $\beta
_{V}=f\left( \beta \right) .$ This implies that a correspondence between the
Gross-Witten model and the Chern-Simons matrix model holds for both the
weak-coupling and the strong-coupling regimes of the model. As seen above
with the Toeplitz determinant representation of the matrix model, and also
in \cite{Szabo:2010qv} using orthogonal polynomials, the weak-coupling limit
follows in a straightforward way. This result, as shown in \cite{Jose:1977gm}%
, also follows by considering decimation \cite{Kadanoff:1976jb}. It is
explained in \cite{Jose:1977gm} that, after a few iterations of the
Kadanoff-Migdal decimation procedure, any interaction function at reasonably
low temperatures generates a new interaction of the Villain type%
\begin{equation*}
\mathrm{e}^{V_{V}\left( \theta -\theta ^{\prime }\right) }=\sum_{m=-\infty
}^{\infty }\mathrm{e}^{-\beta _{V}(\theta -\theta ^{\prime }-2\pi m)^{2}/2}.
\end{equation*}%
In addition, the critical properties of the two models could be identical by
conveniently choosing $\beta _{V}=f\left( \beta \right) $ a function of $%
\beta $ \cite{Jose:1977gm}. In particular, a comparison of the interaction
form for weak and strong coupling showed equivalence if 
\begin{eqnarray}
f\left( \beta \right) &=&\beta \qquad \ \ \ \ \ \ \ \ \ \ \ \ \ \ \ \ \text{%
for\ \ \ }\beta \rightarrow \infty  \notag \\
f\left( \beta \right) &=&\left[ 2\ln \left( 2/\beta \right) \right]
^{-1}\qquad \text{for\ \ \ }\beta \rightarrow 0.  \label{limit}
\end{eqnarray}%
Notice that to compare with the theta function as it appears in the matrix
model (\ref{UCS}), we have to take into account that%
\begin{equation}
\sum_{n=-\infty }^{\infty }\mathrm{e}^{-\beta \left( \theta +2\pi n\right)
^{2}}=\frac{1}{\sqrt{4\pi \beta }}\sum_{n=-\infty }^{\infty }\mathrm{e}%
^{-n^{2}/(4\beta )}\mathrm{e}^{in\theta }.  \label{inv}
\end{equation}%
The r.h.s. of (\ref{inv}) is the series expansion of the theta function in (%
\ref{UCS}). Thus, the decimation of the weight function of the Gross-Witten
model coincides with the Villain approximation and it leads to the weight
function of the unitary Chern-Simons matrix model (\ref{UCS}). In the
weak-coupling limit $\beta \rightarrow \infty $, we obtain that $%
g_{s}=1/\beta $ and hence that $g_{s}=g_{\mathrm{YM}}^{2}/2.$ This also
follows from (\ref{Gaussian}).

The strong-coupling regime $\beta \rightarrow 0$ is specially interesting.
Notice that taking $\beta =0$ ($g_{YM}^{2}=\infty $) directly leads to the
circular ensemble \cite{Mehta}%
\begin{equation*}
Z_{N}^{\mathrm{GW}}(\beta =0)=\int_{0}^{2\pi }\prod\limits_{i<j}\,\sin
^{2}\left( \frac{\theta _{i}-\theta _{j}}{2}\right) \prod\limits_{i=1}^{N}\,%
\mathrm{d}\theta _{i}\ .
\end{equation*}%
In \cite{Szabo:2010qv}, the relationship between this model and Chern-Simons
theory was studied. The result (\ref{limit}) leads to a refined
understanding of this limit $\beta \rightarrow 0.$ In particular, the
relationship between the coupling constants is now logarithmic $g_{s}=2\ln
\left( 2/\beta \right) =2\ln \left( g_{\mathrm{YM}}^{2}\right) $ and hence,
the coupling constant of the Gross-Witten model is related to the $q$
parameter of Chern-Simons theory $q=1/g_{\mathrm{YM}}^{4}$. To summarize,
including the prefactors%
\begin{eqnarray}
\mathrm{e}^{\beta \,\cos \theta } &\approx &\frac{\mathrm{e}^{\beta }}{\sqrt{%
2\pi \beta }}\Theta ({\,\mathrm{e}}\,^{{\,\mathrm{i}\,}\theta _{j}}|{\mathrm{%
e}}\,^{{-1/\beta }}),\text{ with }g_{s}=g_{\mathrm{YM}}^{2}/2\text{ for }g_{%
\mathrm{YM}}\rightarrow 0,  \label{result} \\
\mathrm{e}^{\beta \,\cos \theta } &\approx &\Theta ({\,\mathrm{e}}\,^{{\,%
\mathrm{i}\,}\theta _{j}}|\,{\mathrm{e}}^{{-1/\beta }_{\mathrm{V}}{}}),\text{
with }g_{s}=2\log (g_{\mathrm{YM}}^{2})\text{ for }g_{\mathrm{YM}%
}\rightarrow \infty .  \notag
\end{eqnarray}%
It is also worth mentioning that, already in the original paper on the XY
model \cite{Villain:1974ir}, an approximation valid for both limits was also
given%
\begin{equation}
\mathrm{e}^{\beta \,\cos \theta }\approx R_{V}\left( \beta \right)
\sum_{m=-\infty }^{\infty }\mathrm{e}^{-\beta _{V}(\theta -2\pi m)^{2}/2},
\label{general}
\end{equation}%
with%
\begin{eqnarray}
R_{V}\left( \beta \right)  &=&I_{0}\left( \beta \right) \sqrt{2\pi \beta _{V}%
}  \label{general2} \\
\mathrm{e}^{-\beta _{V}/2} &=&I_{1}\left( \beta \right) /I_{0}(\beta ), 
\notag
\end{eqnarray}%
where $I_{1}\left( \beta \right) $ and $I_{0}(\beta )$ are, as in (\ref%
{Bessel}), Bessel functions. These are the first two Fourier coefficients of
the expansion of the l.h.s. of (\ref{general}). The values for the prefactor 
$R_{V}\left( \beta \right) $ and the coupling constant $\beta _{V}$ are then
found by imposing the first two Fourier coefficients of the two expressions
in (\ref{general}) to coincide. The limits $\beta \rightarrow 0$ and $\beta
\rightarrow \infty $ of (\ref{general2}) coincide with the previous results.
It was shown in \cite{Villain:1974ir} that this approximation is rather
good, even for values close to the critical temperature. This approximation
has been studied in further detail in \cite{Janke:1986ej}, where it was
found that the convergence is better for the strongly coupled regime $\beta
\rightarrow 0.$ In addition, the result (\ref{general}) can be extended,
with good convergence in both limits, to the case where the original
function is $\omega \left( \theta \right) =\beta \cos \theta +\gamma \cos
\left( 2\theta \right) $ \cite{Janke:1986ej}. This suggests that not only
the Gross-Witten model, but more complex multicritical unitary matrix models 
\cite{Periwal}, can also be expressed in terms of the unitary Chern-Simons
matrix model, only with a more sophisticated expression for the coupling
constant and the prefactor.

\subsection{Gaussian behavior and closed string interpretation at weak
coupling}

The relationship between the Gross-Witten model in the weak-coupling limit
and the semiclassical limit of the Chern-Simons matrix model, indicates that
the Gross-Witten model in this regime should be related to closed
topological strings \cite{Ooguri:2002gx}. The reason is that the
semiclassical limit ($k\rightarrow \infty $) of the $U(N)$ Chern-Simons on $%
S^{3}$ free energy also coincides with the nonperturbative part of the total
free energy. The Chern-Simons free energy can be suitably expressed in terms
of nonperturbative and perturbative contributions%
\begin{equation}
F_{\mathrm{CS}}=\log Z_{\mathrm{CS}}=F_{\mathrm{np}}+F_{\mathrm{p}}\ .
\label{np+p}
\end{equation}%
The nonperturbative contribution $F_{\mathrm{np}}$ is the logarithm of the
measure factor in the path integral, which is not captured by Feynman
diagrams, and it gives the exact Chern-Simons partition function in the
semiclassical limit $k\rightarrow \infty $~\cite[eq.~(2.8)]{Ooguri:2002gx}.
It has the explicit expression%
\begin{equation}
F_{\mathrm{np}}=\log \Big(\,\frac{\left( 2\pi \,g_{s}\right) ^{N^{2}/2}}{%
\mathrm{vol}\big(U(N)\big)}\,\Big)\ .  \label{FNP}
\end{equation}%
This free energy (\ref{FNP}), given by a Hermitian Gaussian matrix model,
has an expansion which can be interpreted in terms of closed topological
string theory on the resolved conifold geometry~\cite{Ooguri:2002gx}. 
See~\cite{Ooguri:2002gx} for equivalent string theory and gauge theory
interpretations.

Due to the correspondence between the Chern-Simons and the Gross-Witten
matrix models, the latter should have (\ref{FNP}) as free energy in the
weak-coupling limit $\beta \rightarrow \infty $. Consider the expression for
the free energy of the Gross-Witten model for finite $N$ and $g_{\mathrm{YM}%
}\rightarrow 0$ \cite{Neuberger:1980qh}%
\begin{equation*}
F_{\mathrm{GW}}\simeq \frac{2N}{g_{\mathrm{YM}}^{2}}-\frac{N}{2}\ln (2\pi )-%
\frac{N^{2}}{2}\ln (\frac{2}{g_{\mathrm{YM}}^{2}})+\sum_{j=1}^{N-1}\ln j!+%
\mathcal{O}(g_{YM}^{2}).
\end{equation*}%
This coincides with the free energy of a Hermitian Gaussian matrix model 
\cite{Mehta}%
\begin{equation}
Z_{\mathrm{G}}\left( \beta \right) =\mathrm{e}^{\beta N}\int_{-\infty
}^{\infty }\mathrm{e}^{-\sum_{j=1}^{N}\beta x_{j}^{2}/2}\prod_{j<k}\left(
x_{j}-x_{k}\right) ^{2}.
\end{equation}%
This relationship between the Gross-Witten and the Gaussian matrix model
agrees with \cite{Baik:2007}. Notice that the term $\mathrm{e}^{\beta N}$
actually corresponds to the $\mathrm{e}^{\beta }$ term in the Villain
approximation (\ref{result}) and implies that one has to consider the
Gross-Witten model with the potential (\ref{full-V})\footnote{%
This form of the potential corresponds exactly to the Wilson lattice action.}
, instead of the one in (\ref{GW}).

\section{The Villain lattice action and Abelian/$q$-deformed 2d Yang-Mills
theory}

We begin by discussing the heat-kernel action, which was introduced in
lattice gauge theory, at least in part, as an alternative to the Wilson
action that also provided a natural extension to the non-Abelian case \cite%
{Susskind:1979up,Stone:1978pe,Menotti:1981ry} of the Villain approximation
of the two-dimensional XY model \cite{Villain:1974ir}, which had been
crucially used in the study of $U(1)$ lattice gauge theories \cite%
{Banks:1977cc,Peskin:1977kp}.

The setting is quantum Yang-Mills theory with gauge group $U(N)$ on an
oriented closed Riemann surface $\Sigma _{h}$ of genus $h$ and unit area
form $\mathrm{d}\mu $~ \cite{review}. The action is%
\begin{equation}
S_{\mathrm{YM}}=-\frac{1}{4g_{YM}}\,\int_{\Sigma _{h}}\,\mathrm{d}\mu ~%
\mathrm{Tr}\,F^{2}\ ,  \label{continuum}
\end{equation}%
where $g_{s}$ plays the role of the coupling constant, $F$ is the field
strength of a matrix gauge connection, and $\mathrm{Tr}\,$ is the trace in
the fundamental representation of $U(N)$. A lattice regularization of the
gauge theory relies on a triangulation of the two-dimensional manifold $%
\Sigma $ with group matrices situated along the edges~\cite{Migdal75}. The
path integral is then approximated by the finite-dimensional unitary matrix
integral%
\begin{equation}
\mathcal{Z}_{\mathrm{M}}=\int \,\prod_{\mathrm{edges}~\ell }\,\mathrm{d}%
U_{\ell }~\prod\limits_{\mathrm{plaquettes}~P}\,Z_{P}\left[ U_{P}\right] \ ,
\label{latticeint}
\end{equation}%
where $\mathrm{d}U_{\ell }$ denotes Haar measure on $SU(N)$ and the holonomy 
$U_{P}=\prod\nolimits_{\ell \in P}\,U_{\ell }$ is the ordered product of
group matrices along the links of a given plaquette. The local factor $Z_{P}%
\left[ U_{P}\right] $ is a suitable gauge invariant lattice weight that
converges in the continuum limit to the Boltzmann weight for the Yang-Mills
action (\ref{continuum}). This expression for $\mathcal{Z}_{\mathrm{M}}$
leads to the matrix models presented in the previous section after the
suitable choice of lattice action.

Let us discuss now the use of the heat kernel for the lattice weight $Z_{P}%
\left[ U_{P}\right] $. In this case, the lattice action has many interesting
features~\cite{Menotti:1981ry} and is the usual choice in two-dimensional
Yang-Mills theory~\cite{review}. It leads to the well-known group theory
expansion of the partition function~\cite{Migdal75,Rusakov}%
\begin{equation}
\mathcal{Z}_{\mathrm{M}}=\sum_{\lambda }\,\left( \dim \lambda \right)
^{2-2h}\,\exp \big(-g_{YM}^{2}\,C_{2}(\lambda )\big)\ ,  \label{HK1}
\end{equation}%
where the sum runs through all isomorphism classes $\lambda $ of irreducible
representations of the $SU(N)$ gauge group, $\dim \lambda $ is the dimension
of the representation $\lambda $, and $C_{2}(\lambda )$ is the quadratic
Casimir invariant of $\lambda $. This expression for the partition function
is a particular case of the propagator \cite{Menotti:1981ry,review}%
\begin{equation*}
\exp \left( -S_{\mathrm{HK}}\left( U\right) \right) =\left\langle \mathbb{I}%
\right\vert \exp \left( \frac{1}{2}g_{YM}^{2}\Delta \right) \left\vert
U\right\rangle \equiv K(U,\frac{g_{YM}^{2}}{2}),
\end{equation*}%
\newline
which can be written as \cite{Menotti:1981ry}%
\begin{equation}
K(U,\frac{g_{YM}^{2}}{2})=\sum_{\lambda }\dim \lambda \chi _{\lambda }\left(
U\right) \exp \left( -g_{YM}^{2}C_{2}(\lambda \right) ),  \label{K}
\end{equation}%
\newline
where the sum runs over all irreducible unitary representations of the gauge
group, $\chi _{\lambda }\left( U\right) $ is the character of such a
representation, $\dim \lambda =$ $\chi _{\lambda }\left( \mathbb{I}\right) $
its dimension and $C_{2}(\lambda )$ the Casimir of the representation.
Writing (\ref{K}) in terms of the elements of the Young tableaux that labels
the representation $\lambda $ leads to a discrete matrix model
representation \cite{review}.

\subsubsection{$q$ deformation}

In recent years, the relationship between two-dimensional Yang-Mills theory
and Chern-Simons theory on Seifert manifolds has been understood in further
detail \cite{Aganagic:2004js,Beasley:2005vf,Beasley:2009mb} (see also \cite%
{Blau:2006gh,Arsiwalla:2005jb,Caporaso:2005ta}). Seifert manifolds $M(h,p)$
are nontrivial circle bundles (of monopole degree $p$) over two-dimensional
surfaces of genus $h$. The simplest case, the trivial fibration, $%
M(h,0)=\Sigma _{h}\times S^{1}$ was studied in detail in \cite%
{Witten:1988hf,Blau:1993tv}.

Chern-Simons theory on Seifert manifolds has been the subject of much
interest in the study of topological strings \cite{Marino:2004uf} and has a
direct relationship with a $q$ deformation of the two-dimensional Yang-Mills
theory discussed above, when the manifold is a sphere, $\Sigma _{0}=S^{2}$.
In particular, the partition function of $q$-deformed 2d YM on a closed
Riemann surface of genus $h$ is given by \cite{Aganagic:2004js}%
\begin{equation}
Z_{YM}^{q}(\Sigma _{h})=\sum_{\lambda }\left( \dim _{q}\left( \lambda
\right) \right) ^{2-2h}q^{\frac{p}{2}C_{2}(\lambda )},  \label{q}
\end{equation}%
where $\dim _{q}\left( \lambda \right) $ is the $q$ deformation of the
dimensions of $sl_{n}$ representations, i.e. the quantum dimensions $\dim
_{q}(\lambda )$~\cite{fuchs}, $p$ is a positive integer parameter and, as
usual, $C_{2}(\lambda )$ is the Casimir of the representation $\lambda $.
This is related to the partition function of Chern-Simons theory on a circle
fibration (with Chern class $p$) over $\Sigma _{h}$, which is a Seifert
space. In the case $\Sigma _{h}=S^{2}$ and $p>1$, the Seifert manifold is
the lens space $S^{3}/\mathbb{Z}_{p}.$ If $p=1,$ then the connection is with
Chern-Simons theory on $S^{3}$, the case studied here.

\subsection{Abelianization and q deformation}

The propagator (\ref{K}) can be alternatively written in terms of the
elements of $U(N)$, and then one obtains a unitary matrix model expression.
When written in terms of the invariant angles of the gauge group, it is
given by \cite{Menotti:1981ry}%
\begin{equation}
\exp \left( -S_{\mathrm{HK}}\left( \theta _{1},...\theta _{N}\right) \right)
=\mathcal{N}\sum_{\left\{ l\right\} =-\infty }^{\infty }\prod\limits_{i<j}%
\frac{\theta _{i}-\theta _{j}+2\pi \left( l_{i}-l_{j}\right) }{2\sin \left[
\theta _{i}-\theta _{j}+2\pi \left( l_{i}-l_{j}\right) \right] }\exp \left[ -%
\frac{1}{g_{YM}^{2}}\sum_{j=1}^{N}\left( \theta _{j}+2\pi l_{j}\right) ^{2}%
\right] ,  \label{HK}
\end{equation}%
where $\mathcal{N}$ stands for some normalization. This is a rather complex
model, and therefore the heat-kernel case, in contrast to the two cases
studied in the previous section, is not studied with a unitary matrix model
but rather with a discrete matrix model that follows from (\ref{HK1}). We
show now that a simplification of (\ref{HK}) leads to the Chern-Simons
matrix model. The exponential part in the r.h.s. is the Abelian $U(1)$
Villain action, which is just a theta function (\ref{Villain})%
\begin{equation}
\exp (-S_{\mathrm{V}}\left( \theta \right) )=\sum_{l=-\infty }^{\infty }%
\mathrm{e}^{-\frac{1}{g_{YM}^{2}}\left( \theta +2\pi l\right) ^{2}},
\label{A-Villain2}
\end{equation}%
and the straightforward generalization to $U(N)$ gives a propagator%
\begin{eqnarray}
\overline{K}(U,\frac{g_{YM}^{2}}{2}) &=&\sum_{\{l\}=-\infty }^{\infty }\exp %
\left[ -\frac{1}{g_{YM}^{2}}\sum_{j=1}^{N}\left( \theta _{j}+2\pi
l_{j}\right) ^{2}\right]   \label{V-action} \\
&=&\prod\limits_{i=1}^{N}\exp (-S_{\mathrm{V}}(\theta _{i}))=\exp \left( -S_{%
\mathrm{Villain}}\left( \theta _{1},...\theta _{N}\right) \right)   \notag
\end{eqnarray}%
that defines an action which is just the direct product of the Abelian
Villain action (\ref{A-Villain2}). The corresponding one-plaquette model is
given by a $U(N)$ matrix integral, which is the unitary integration of the
propagator. The partition function is then given by the matrix model%
\begin{equation}
Z_{N}=\int_{0}^{2\pi }\prod\limits_{i=1}^{N}\,\frac{\mathrm{d}\theta _{i}}{%
2\pi }~\sum_{n=-\infty }^{\infty }\mathrm{e}^{-\frac{1}{g_{YM}^{2}}\left(
\theta _{i}+2\pi n\right) ^{2}}\prod\limits_{k<l}\,\big\vert{\,\mathrm{e}}%
\,^{{\,\mathrm{i}\,}\theta _{k}}-{\,\mathrm{e}}\,^{{\,\mathrm{i}\,}\theta
_{l}}\big\vert^{2},  \label{UCSb}
\end{equation}%
which, recalling the identity (\ref{inv}), is the unitary matrix model
description of $U(N)$ Chern-Simons theory on $S^{3}$ (\ref{UCS}) discussed
so far and previously studied in \cite%
{Okuda:2004mb,Szabo:2010sd,Ooguri:2010yk}. Hence, $Z_{N}=Z_{\mathrm{CS}%
}^{U(N)}\left( S^{3}\right) $, by identifying $g_{s}=g_{YM}^{2}/2$. Thus,
the Abelianization of the heat kernel (\ref{HK}), given by the
straightforward extension of the $U(1)$ Abelian Villain action to the $%
U(1)^{N}$ case (\ref{V-action}), leads to $U(N)$ Chern-Simons theory on $%
S^{3}$.

At the level of the matrix model, this straightforward generalization of the
Villain action was already considered in an interesting paper by Onofri \cite%
{Onofri:1981qk}, that analyzed this problem and presented a detailed study
of the matrix model (\ref{UCSb}). It is remarkable that the results in that
paper describe pure Chern-Simons theory on $S^{3}$, with $U(N)$ and $SU(N)$,
a fact that has not been pointed out so far. Some expressions in \cite%
{Onofri:1981qk} are manifestly Chern-Simons observables, like the partition
function of $U(N)$ Chern-Simons theory on $S^{3}$.

We have seen the Abelianization of the heat kernel, using its expression in
terms of the invariant angles of the gauge group (\ref{HK}). Let us now look
at the relationship between the choice of the Villain lattice action and the 
$q$ deformation of the heat-kernel 2d Yang-Mills theory, using instead the
character expansion (\ref{K}).

Notice that the $q$-deformed 2d Yang-Mills (\ref{q}) is just given by the
usual expression for two-dimensional Yang-Mills theory based on the
heat-kernel action and defined on a compact manifold of genus $h$, but with $%
q$ dimensions instead of ordinary dimensions, $\dim (\lambda )\rightarrow
\dim _{q}(\lambda )$. We will now show that choosing (\ref{V-action}) with
the Villain action (\ref{A-Villain}) as a lattice action is actually
equivalent to $q$-deformed 2d Yang-Mills. Since choosing the Villain action
is tantamount to an Abelian projection of the two-dimensional Yang-Mills
theory with the heat-kernel lattice action, this implies that the
Abelianization and the $q$ deformation of the latter lead to the same theory.

An expression of the type (\ref{q}) is valid for 2d Yang-Mills on a compact
manifold of genus $h$ and it can be constructed from the central heat
kernel, which is also the propagator on a cylinder \cite{review}%
\begin{equation}
K(\frac{g_{YM}^{2}}{2},U,U^{\prime })=\sum_{\lambda }\chi _{\lambda }\left(
U\right) \chi _{\lambda }\left( U^{\prime }\right) \exp \left(
-g_{YM}^{2}C_{2}(\lambda \right) ),  \label{hk}
\end{equation}%
with holonomies $U$ and $U^{\prime }$ on the two disks of the cylinder and
where $\chi _{\lambda }\left( U\right) $ is the character of the irreducible
representation $\lambda ,$ as in (\ref{K}) above. Recall that $\chi
_{\lambda }\left( U=\mathbb{I}\right) =\dim (\lambda )$. Hence, if $%
U=U^{\prime }=\mathbb{I}$, (\ref{hk})\ is the expression for the partition
function of two-dimensional Yang-Mills theory on $S^{2}$. If only $U^{\prime
}=\mathbb{I},$ then it describes two-dimensional Yang-Mills on a disk%
\begin{equation}
K(\frac{g_{YM}^{2}}{2},U,\mathbb{I})=\sum_{\lambda }\chi _{\lambda }\left(
U\right) \dim \lambda \exp \left( -g_{YM}^{2}C_{2}(\lambda \right) ).
\label{disk}
\end{equation}%
This is the heat kernel and it is also the amplitude of a plaquette that
leads, by gluing, to (\ref{HK}) \cite{review}. Since (\ref{disk}) is also
the fundamental solution of the diffusion equation, then, as is well-known,
two-dimensional Yang-Mills theory based on the heat kernel can be understood
as a diffusion process on the gauge group manifold.

The relationship between the Chern-Simons model constructed with the Villain
action and the $q$ deformation of two-dimensional Yang-Mills theory can be
readily seen by considering the character expansion of the $U(N)$ Villain
action (\ref{V-action}) given in \cite{Onofri:1981qk}%
\begin{equation}
\frac{\mathrm{e}^{-S_{\mathrm{Villain}}(U)}}{Z}=\sum_{m_{1}\geq m_{2}\geq
\ldots \geq m_{N}=0}\chi _{(m_{1},\ldots ,m_{N})}(U)q^{\frac{1}{2}%
\sum_{i=1}^{N}m_{i}^{2}}\prod_{j>i}\left( \frac{q^{m_{i}-m_{j}+j-i}-1}{%
q^{j-i}-1}\right) ,  \label{ch}
\end{equation}%
where the sum is over the integers $\left\{ m_{i}\right\} $ with $i=1,...,N$
and $Z$ is the partition function of the unitary matrix model, which has the
explicit form%
\begin{equation*}
Z=\left( \frac{g_{\mathrm{YM}}^{2}}{8\pi }\right) ^{\frac{N}{2}%
}\prod_{k=1}^{N-1}\left( 1-q^{N-k}\right) ^{k},\qquad q=\mathrm{e}^{-\frac{%
g_{\mathrm{YM}}^{2}}{2}}.
\end{equation*}%
This character expansion of the Villain action suggests that such an action
leads to the usual propagator of two-dimensional Yang-Mills theory based on
the heat kernel (\ref{disk}) but with $q$ dimensions instead of dimensions,
since the last term in (\ref{ch}) gives an explicit expression for quantum
dimensions. An elementary manipulation of the product shows this explicitly%
\begin{equation*}
\prod_{j>i}\left( \frac{q^{m_{i}-m_{j}+j-i}-1}{q^{j-i}-1}\right)
=\prod_{j>i}q^{\frac{m_{j}-m_{i}}{2}}\frac{[{m_{i}-m_{j}+j-i}]_{q}}{[j-i]_{q}%
}=q^{\frac{1}{2}\sum_{l=1}^{N}(N-2l+1)m_{l}}\mathrm{dim}_{q}(\lambda ),
\end{equation*}%
where $\lambda $ denotes the unitary irreducible representation of the gauge
group, characterized by a partition whose Young tableaux has columns with $%
\left\{ m_{i}\right\} $ boxes. The explicit expression for the quantum
dimensions is given, as in \cite{Aganagic:2004js}, by%
\begin{equation*}
\mathrm{dim}_{q}(\lambda )=\prod_{j>i}\frac{[{m_{i}-m_{j}+j-i}]_{q}}{%
[j-i]_{q}},\text{ with the }q\text{-number }[x]_{q}=\frac{q^{\frac{x}{2}%
}-q^{-\frac{x}{2}}}{q^{\frac{1}{2}}-q^{-\frac{1}{2}}}.
\end{equation*}%
Putting all together, we see that the Casimir term is manifest and appears
exactly as in \cite{Aganagic:2004js}%
\begin{equation*}
C_{2}(\lambda
)=\sum_{i=1}^{N}m_{i}^{2}+(N-2i+1)m_{i}=\sum_{i=1}^{N}m_{i}(m_{i}-2i+1)+N%
\sum_{i=1}^{N}m_{i},
\end{equation*}%
and hence we obtain%
\begin{equation}
\frac{\mathrm{e}^{-S_{\mathrm{Villain}}(U)}}{Z}=\sum_{\lambda }\chi
_{\lambda }(U)q^{\frac{1}{2}C_{2}(\lambda )}\mathrm{dim}_{q}(\lambda ).
\label{propagator}
\end{equation}%
Thus, the r.h.s of (\ref{ch}) is the disk amplitude for $q$-deformed
two-dimensional Yang-Mills theory with $p=1$, which, by the same procedure
discussed above, and explained in detail in \cite{review}, leads to the
expression for the partition function (\ref{q}). This identity is actually a
particular case of the Kostant identity, which gives a character expansion
of theta functions of a lattice \cite{Kostant}. This specific form of the
identity was then rediscovered later on, in \cite{Onofri:1981qk}, in the
context of the Villain lattice action, and it also appears much later in 
\cite{Aganagic:2005dh}, where, working from the r.h.s. of (\ref{propagator})
a theta function expression was found, which is, as we have seen here, the
Villain lattice action. In this paper, we have also shown, by using the
equivalent formulation of the lattice action in terms of the invariant
angles (\ref{HK}), that it follows from taking only the Abelian part of the
heat kernel.

The fact that the Abelianization and the $q$ deformation described here are
equivalent is qualitatively consistent with the fact, explained for example
in \cite{Witten:1989rw}, that a $q$ deformation of a Lie group $G$ is not a
group and lacks its symmetry, whereas the maximal torus $T$ of $G$ remains
an ordinary symmetry group after the symmetry breaking inherent in the
transformation of a Lie group into a quantum group.

To conclude, let us mention that inspection of the analogous expression for $%
SU(N)$ in \cite{Onofri:1981qk} shows that $\dim _{q}\lambda $ can also be
written as the character $\chi _{\lambda }\left( T_{q}\right) $, where $%
T_{q}\in SL\left( N,\mathbb{C}\right) $ is given by $\mathrm{diag}\left[
q^{N-1},q^{N-3},...,q^{3-N},q^{1-N}\right] .$ Hence, the propagator is now
of the type $K(\frac{g^{2}}{2},U,T_{q})$ instead of (\ref{disk}), which
suggests that diffusion does not take place in the whole gauge group.
Indeed, the explicit form of the matrix model indicates diffusion on the
maximal torus $U(1)\times ...\times U(1)$ of the gauge group $U(N)$. This is
in agreement with a previous result that related Chern-Simons theory on $%
S^{3}$ with Brownian motion on the Weyl chamber of the gauge group \cite{dHT}%
.

\section{Conclusions and Outlook}

We have seen how the unitary matrix model that describes $U(N)$\
Chern-Simons theory on $S^{3}$ arises from studying two-dimensional
Yang-Mills theory with the Villain lattice action and we have compared it
both with the Wilson and the heat-kernel lattice action cases.

Regarding the former, we have seen that the Gross-Witten model is related to
the Chern-Simons matrix model both in the weak-coupling and the
strong-coupling regimes. As we have seen in Sec. 2.1., one of the
implications is that the Gross-Witten model, which describes two-dimensional
Yang-Mills theory on $\mathbb{R}^{2}$, coincides in the weak-coupling limit
with the nonperturbative part of Chern-Simons theory on $S^{3}$, and
consequently has the same string theory interpretation \cite{Ooguri:2002gx}.
In both cases, the free energy is given by a Hermitian Gaussian matrix
model. In spite of the apparent simplicity of such a matrix model, it is
actually relevant in the study of subsectors of $\mathcal{N}$=4
supersymmetric gauge theory and their relationship with two-dimensional
Yang-Mills theory (see \cite{Giombi:2009ms}, for example). It is possible
that taking into account Wilson loops in our discussion would lead to a
relationship with that line of research.

Notice also that the general approximation of the Gross-Witten weight (\ref%
{general}) implies that a small modification of the Gross-Witten model
potential leads to the Chern-Simons matrix model. The interest of this
result lies in a possible connection between the Chern-Simons matrix model
and the unitary matrix models that appear in the study of phase transitions
of weakly coupled gauge theories \cite{Aharony:2003sx}.

The heat-kernel has a character expansion which is the basis of the study of
2d Yang-Mills theory. However, it also can be expressed in the invariant
angles of the gauge group (the unitary group in our case), as pointed out in 
\cite{Menotti:1981ry}. The unitary matrix model that follows from this
representation is not used in the heat-kernel case, due to its complexity.
However, we also have seen that an Abelian projection of the heat-kernel
lattice action leads to a $U(1)^{N}$ lattice action, which is the Villain
lattice action. After unitary integration of the resulting propagator, the
corresponding matrix model is now the $U(N)$ Chern-Simons matrix model for $%
S^{3}$. On the other hand, since the character expansion of the Villain
lattice action gives the $q$-deformed 2d Yang-Mills propagator
(Kostant-Onofri identity), we see that the Abelianization of (\ref{HK})
coincides with the $q$ deformation of (\ref{disk}), given by (\ref%
{propagator}). In addition, it shows how the $q$ propagator directly leads
to the unitary Chern-Simons matrix model, instead of the (equivalent)
Chern-Simons Hermitian matrix model.

Precisely, and to conclude, in the Appendix, the relationship between the
unitary and the Hermitian versions of the Chern-Simons matrix model is
studied in detail, focusing also in the rotation of the contours of
integration.

\subsection*{Acknowledgments}

Thanks to Sergio Iguri and Fokko van de Bult for comments and
correspondence. The work of MT has been supported by the project
\textquotedblleft Probabilistic approach to finite and infinite dimensional
dynamical systems\textquotedblright\ (PTDC/MAT/104173/2008) at the
Universidade de Lisboa.

\newpage

\appendix

\section{Unitary vs Hermitian and integration contours in matrix models}

In this paper we have focused on the unitary matrix model that describes
Chern-Simons theory on $S^{3}$. It was first considered in \cite%
{Okuda:2004mb} although we have seen that it was already studied in detail
in \cite{Onofri:1981qk}. In contrast to the Hermitian matrix model, no
additional work has been done using the unitary model, with the exception of
its recent appearance in the study of matrix models in Donaldson-Thomas
theory \cite{Ooguri:2010yk,Szabo:2010sd}. The Hermitian matrix model, in the 
$S^{3}$ case reads \cite{Marino:2002fk} 
\begin{equation}
Z_{\mathrm{CS}}^{U(N)}\left( S^{3}\right) =\int_{-\infty }^{\infty }\mathrm{e%
}^{-\sum_{j=1}^{N}x_{j}^{2}/\left( 2g_{s}\right) }\prod_{j<k}\left( 2\sinh 
\frac{\pi (x_{j}-x_{k})}{2}\right) ^{2}\prod_{j=1}^{N}\frac{\mathrm{d}x_{j}}{%
2\pi }.  \label{M}
\end{equation}%
Both models, the unitary (\ref{UCS}) and the Hermitian (\ref{M}), are solved
with orthogonal polynomials that have the same orthogonality properties, and
hence the corresponding observables of the matrix model, like the partition
function for example, coincide. These polynomials are the Stieltjes-Wigert
polynomials in the Hermitian case \cite{Tierz:2002jj} and the Rogers-Szeg%
\"{o} polynomials in the case of the unit circle \cite%
{Dolivet:2006ii,Szabo:2010qv}. The connection between these two systems of
orthogonal polynomials allows to explain the relationship between both
matrix models, as was shown in \cite{Dolivet:2006ii}. However, it is
interesting to have a more immediate relationship between the models. Let us
write the model (\ref{M}) in its trigonometric version: 
\begin{equation}
\widetilde{Z}_{\mathrm{CS}}^{U(N)}\left( S^{3}\right) =\int_{-\infty
}^{\infty }\mathrm{e}^{-\frac{1}{2g_{s}}\sum_{j=1}^{N}u_{j}^{2}}\prod_{j<k}%
\left( 2\sin \frac{u_{j}-u_{k}}{2}\right) ^{2}\prod_{j=1}^{N}\frac{du_{j}}{%
2\pi }.  \label{trig}
\end{equation}%
We will now show its relationship with (\ref{M}). The first step is to
relate (\ref{trig}) with the unitary model (\ref{UCS}), by transforming the
Vandermonde determinant%
\begin{equation}
\prod_{j<k}\left( 2\sin \frac{u_{j}-u_{k}}{2}\right) ^{2}=\prod_{j<k}|%
\mathrm{e}^{iu_{j}}-\mathrm{e}^{iu_{k}}|^{2},
\end{equation}%
and the weight function, making the range of integration compact and using
also the identity (\ref{inv}) 
\begin{eqnarray}
\int_{-\infty }^{\infty }\prod_{j<k}|\mathrm{e}^{iu_{j}}-\mathrm{e}%
^{iu_{k}}|^{2}\mathrm{e}^{-\frac{1}{2g_{s}}\sum_{j=1}^{N}u_{j}^{2}}%
\prod_{j=1}^{N}\frac{du_{j}}{2\pi } &=&\frac{g_{s}^{\frac{N}{2}}}{(2\pi )^{%
\frac{N}{2}}}\int_{0}^{2\pi }\prod_{j=1}^{N}\frac{du_{j}}{2\pi }%
\sum_{n=-\infty }^{\infty }\mathrm{e}^{-\frac{n^{2}g_{s}}{2}%
+inu_{j}}\prod_{j<k}|\mathrm{e}^{iu_{j}}-\mathrm{e}^{iu_{k}}|^{2}  \notag \\
&=&\frac{g_{s}^{\frac{N}{2}}}{(2\pi )^{\frac{N}{2}}}\int_{0}^{2\pi
}\prod_{j=1}^{N}\frac{du_{j}}{2\pi }\Theta (q|u_{j})\prod_{j<k}|\mathrm{e}%
^{iu_{j}}-\mathrm{e}^{iu_{k}}|^{2}.
\end{eqnarray}%
Hence, we see that the trigonometric matrix model (\ref{trig}) is equivalent
to the unitary Chern-Simons matrix model (\ref{UCS}). It is also related to (%
\ref{M}) by the change of variables $x=iu$ 
\begin{eqnarray}
&&\int_{-\infty }^{\infty }\mathrm{e}^{-\sum_{j=1}^{N}x_{j}^{2}/\left(
2g_{s}\right) }\prod_{j<k}\left( 2\sinh \left( \frac{\pi (x_{j}-x_{k})}{2}%
\right) \right) ^{2}\prod_{j=1}^{N}dx_{j}  \label{hyptrig} \\
&=&\mathrm{e}^{\frac{i\pi N(N+1)}{4}}\int_{i\infty }^{-i\infty }\mathrm{e}%
^{\sum_{j=1}^{N}u_{j}^{2}/\left( 2g_{s}\right) }\prod_{j<k}\left( 2\sin
\left( \frac{\pi (u_{j}-u_{k})}{2}\right) \right) ^{2}\prod_{j=1}^{N}du_{j}.
\notag
\end{eqnarray}%
However, notice that the integral in the trigonometric model (\ref{hyptrig})
is actually over the imaginary line of the complex plane. Then, to show the
equivalence of (\ref{M}) and the unitary model (\ref{UCS}), the r.h.s of (%
\ref{hyptrig}) has to be equal to (\ref{trig}). What we get now is (\ref%
{trig}) but with the opposite sign in the weight and a different integration
contour. If we were able to rotate the contour to the real axis, then the
sign in the exponential of the weight can be corrected just by complex
conjugation of (\ref{hyptrig}), recalling $g_{s}$ as a purely imaginary
quantity. Hence, one needs to rotate the contour to the real line and due to
the results above mentioned, we know that this has to be the case.

This rotation of the contour is also directly related to the fact that the
actual derivation of the matrix model for Seifert manifolds leads to
integral expressions with contours on the complex plane \cite%
{LR,Marino:2002fk}, and it is just assumed that they can be rotated into the
real axis \cite{Marino:2002fk}, leading then to expressions such as (\ref{M}%
).

It is then worthwhile to examine the rotation of contours more carefully.
For this, we use certain particular cases of multivariate hyperbolic
hypergeometric integrals, studied in \cite{Fokko}, that we can easily
identify with the Chern-Simons matrix integrals but with complex integration
contours. The integrals in \cite{Fokko} are written in terms of the
hyperbolic Gamma function. However, due to the following property of the
hyperbolic Gamma function \cite{Fokko} 
\begin{equation}
\frac{1}{\Gamma _{h}(z|\omega _{1},\omega _{2})\Gamma _{h}(-z|\omega
_{1},\omega _{2})}=-4\sin \left( \frac{\pi z}{\omega _{1}}\right) \sin
\left( \frac{\pi z}{\omega _{2}}\right) ,  \label{g-identity}
\end{equation}%
it is then immediate to identify such integrals in \cite{Fokko} with those
given by the matrix model description of Chern-Simons theory on $S^{3}$. To
see that this is the case, let us first give some definitions in \cite{Fokko}
and then we quote Proposition 5.3.19 in \cite{Fokko}.\newline
\newline
\textbf{Definition (hook)}:\emph{\ A hook $W_{\phi _{1},\phi _{2}}$ is a
contour parametrized by $W_{\phi _{1},\phi _{2}}(s)=s\mathrm{e}^{i\phi _{1}}$
for $s\in (-\infty ,0]$ and $W_{\phi _{1},\phi _{2}}(s)=s\mathrm{e}^{i\phi
_{2}}$ for $s\in \lbrack 0,\infty )$. The following quantities are defined
as well $\phi _{+}=\max (\arg (\omega _{1}),\arg (\omega _{2}))$ and $\phi
_{-}=\min (\arg (\omega _{1}),\arg (\omega _{2}))$ and, in terms of $\phi
_{+}$ and $\phi _{-},$ the domain $\mathcal{A}_{+}$ in the complex plane}%
\begin{equation*}
\mathcal{A}_{+}=(\phi _{+}-\pi ,\frac{\phi _{+}+\phi _{-}-\pi }{2}).
\end{equation*}%
After using (\ref{g-identity}), Proposition 5.3.19 reads \cite{Fokko} 
\newline
\newline
\textbf{Proposition 1}: For $t\in \mathbb{Z}_{>0}$ we have 
\begin{eqnarray}
J_{N,t}(\omega _{1},\omega _{2}) &\equiv &\frac{1}{(\sqrt{-\omega _{1}\omega
_{2}})^{N}N!}\int_{C}\mathrm{e}^{\frac{t\pi i}{2\omega _{1}\omega _{2}}%
\sum_{j=1}^{N}x_{j}^{2}}\prod_{j<k}(-4)\sin \left( \frac{\pi (x_{j}-x_{k})}{%
\omega _{1}}\right) \sin \left( \frac{\pi (x_{j}-x_{k})}{\omega _{2}}\right)
\prod_{j}dx_{j}  \notag \\
&=&\frac{\mathrm{e}^{-\frac{i\pi N^{2}}{4}}2^{\frac{N}{2}}}{t^{\frac{N}{2}}}%
\mathrm{e}^{-\frac{\pi iN^{2}(N-1)(\omega _{1}^{2}+\omega _{2}^{2})}{%
6t\omega _{1}\omega _{2}}}\prod_{j=1}^{N}\left( 2\sin \left( \frac{2\pi j}{t}%
\right) \right) ^{N-j}
\end{eqnarray}%
where the contour of integration $C$ is a hook $W_{\phi _{1},\phi _{2}}$
with $\phi _{1},\phi _{2}\in \mathcal{A}_{+}$.\newline

This result can be extended to $t\in \mathbb{Z}_{<0}$ by using a property of
invariance under complex conjugation \cite{Fokko} 
\begin{equation}
J_{N,t}(\omega _{1},\omega _{2})=\overline{J_{N,-t}(-\overline{\omega }_{1},-%
\overline{\omega }_{2})}.  \label{compxconj}
\end{equation}%
To compare with the expression given by the Chern-Simons matrix model, we
take $\omega _{1}=\omega _{2}=i$; therefore $\mathcal{A}_{+}=\left( -\frac{%
\pi }{2},0\right) $. We also perform the change of variables $2\pi
x_{i}=u_{i}$ and identify $t=2(k+N)$. Then the integral can be written as 
\begin{equation}
\int_{C}\prod_{j<k}4\sinh ^{2}(\frac{u_{j}-u_{k}}{2})\mathrm{e}^{\frac{1}{%
2g_{s}}\sum_{j}u_{j}^{2}}\prod_{j}\frac{du_{j}}{2\pi }=N!\mathrm{e}^{-\frac{%
i\pi N^{2}}{4}}(k+N)^{-\frac{N}{2}}\mathrm{e}^{-\frac{g_{s}N^{2}(N-1)}{12}%
}\prod_{j=1}^{N}\left( 2\sin \left( \frac{\pi j}{k+N}\right) \right) ^{N-j},
\label{inta}
\end{equation}%
or, equivalently [using (\ref{compxconj})] 
\begin{equation}
\int_{C}\prod_{j<k}4\sinh ^{2}(\frac{u_{j}-u_{k}}{2})\mathrm{e}^{-\frac{1}{%
2g_{s}}\sum_{j}u_{j}^{2}}\prod_{j}\frac{du_{j}}{2\pi }=N!\mathrm{e}^{\frac{%
i\pi N^{2}}{4}}(k+N)^{-\frac{N}{2}}\mathrm{e}^{\frac{g_{s}N^{2}(N-1)}{12}%
}\prod_{j=1}^{N}\left( 2\sin \left( \frac{\pi j}{k+N}\right) \right) ^{N-j}.
\label{intb}
\end{equation}%
If we move the contour of integration of (\ref{intb}) to the real axis, we
get the expression for the Chern-Simons partition function computed in \cite%
{Tierz:2002jj} using the Hermitian matrix model. Notice that in \cite{LR},
when restricted to the case $S^{3}$, the partition function is then given by
a single complex integral over the line $\mathbb{R}\times \mathrm{e}^{-\frac{%
i\pi }{4}}$, which is exactly (\ref{intb}) for that choice of contour.

We can use (\ref{inta}) and (\ref{intb}) to establish the relation between
the unitary and Hermitian matrix models. If we rotate the contour of (\ref%
{inta}) to the imaginary axis [that means, taking the hook $W_{-\frac{\pi }{2%
},-\frac{\pi }{2}}(s)$], we get the unitary matrix model with $q=\mathrm{e}%
^{-g_{s}}$, by using the identities previously derived 
\begin{eqnarray}
&&\int_{-\infty }^{\infty }\prod_{j<k}(-4)\sin ^{2}(\frac{u_{j}-u_{k}}{2})%
\mathrm{e}^{-\frac{1}{2g_{s}}\sum_{j=1}^{N}u_{j}^{2}}\prod_{j}(-i)\frac{%
du_{j}}{2\pi }  \notag \\
&=&\mathrm{e}^{\frac{i\pi N(N+2)}{2}}\int_{-\infty }^{\infty
}\prod_{j<k}|e^{iu_{j}}-e^{iu_{k}}|^{2}\mathrm{e}^{-\frac{1}{2g_{s}}%
\sum_{j=1}^{N}u_{j}^{2}}\prod_{j}\frac{du_{j}}{2\pi }  \notag \\
&=&N!\mathrm{e}^{-\frac{i\pi N^{2}}{4}}(k+N)^{-\frac{N}{2}}\mathrm{e}^{-%
\frac{g_{s}N^{2}(N-1)}{12}}\prod_{j=1}^{N}\left( 2\sin \left( \frac{\pi j}{%
k+N}\right) \right) ^{N-j}.
\end{eqnarray}%
By noting that (\ref{intb}) is precisely (\ref{M}), then 
\begin{eqnarray*}
Z &=&\int_{0}^{2\pi }\prod_{j}\frac{du_{j}}{2\pi }\Theta
(q|u_{j})\prod_{j<k}|\mathrm{e}^{iu_{j}}-\mathrm{e}^{iu_{k}}|^{2}=\left( 
\frac{2\pi }{g_{s}}\right) ^{\frac{N}{2}}\mathrm{e}^{-i\pi N(N+1)}\mathrm{e}%
^{-\frac{g_{s}N^{2}(N-1)}{6}}Z_{CS}^{U(N)}(S^{3}) \\
&=&\mathrm{e}^{-i\pi N(N+1)}N!\prod\limits_{i=1}^{N-1}(1-q^{i})^{N-i}.
\end{eqnarray*}%
This establishes the precise relation between the unitary and Hermitian
Chern-Simons matrix models, derived directly from contour rotation and is
our final result. It is worth mentioning that other integrals considered in 
\cite{Fokko} correspond to the Chern-Simons matrix model on $S^{3}$ but for
other gauge groups . For example, Proposition 5.3.18 in \cite{Fokko} reads 
\newline
\newline
\textbf{Proposition 2}: For $t\in \mathbb{Z}_{>0}$ we have 
\begin{eqnarray}
\widetilde{J}_{N,t}(\omega _{1},\omega _{2}) &\equiv &\frac{1}{(\sqrt{%
-\omega _{1}\omega _{2}})^{N}N!}\int_{C}\prod_{j}\mathrm{e}^{\frac{t\pi i}{%
\omega _{1}\omega _{2}}x_{j}^{2}}dx_{j}\prod_{j<k}4\sin (\frac{\pi
(x_{j}-x_{k})}{\omega _{1}})\sin (\frac{\pi (x_{j}-x_{k})}{\omega _{2}}) 
\notag \\
&&\times \prod_{l<r}4\sin (\frac{\pi (x_{l}+x_{r})}{\omega _{1}})\sin \left( 
\frac{\pi (x_{l}+x_{r})}{\omega _{2}}\right) \prod_{j=1}^{N}(-1)4\sin \left( 
\frac{\pi x_{j}}{\omega _{1}}\right) \sin \left( \frac{\pi x_{j}}{\omega _{2}%
}\right)  \\
&=&\frac{\mathrm{e}^{-\frac{i3\pi N}{4}}}{t^{\frac{N}{2}}}\mathrm{e}^{-\frac{%
\pi iN(N+1)(2N+1)(\omega _{1}^{2}+\omega _{2}^{2})}{6t\omega _{1}\omega _{2}}%
}\prod_{j<k}4\sin \left( \frac{\pi (j+k)}{t}\right) \sin \left( \frac{\pi
(j+k)}{t}\right) \prod_{j=1}^{N}2\sin \left( \frac{2\pi j}{t}\right)   \notag
\end{eqnarray}%
where the contour of integration $C$ is a hook $W_{\phi _{1},\phi _{2}}$
with $\phi _{1},\phi _{2}\in \mathcal{A}_{+}$. This gives an evaluation for
the partition function of the Chern-Simons matrix model for $S^{3}$ and the
symplectic gauge group $Sp(2N)$.\newline

\end{document}